\documentstyle[twocolumn,prl,aps,floats,psfig,amsmath,epsfig]{revtex}

\newcommand{\ApJL}{Astrophys. J. Lett.}
\newcommand{\ApJ}{Astrophys. J.}
\newcommand{\PRL}{Phys. Rev. Lett.}
\newcommand{\PRD}{Phys. Rev. D}

\newcommand{\cmb}{\Theta}

\newcommand{\bfr}{{\mathbf{r}}}

\newlength{\tskip}
\newlength{\colwidth}

\newcommand{\beq}{\begin{equation}}
\newcommand{\eeq}{\end{equation}}
\newcommand{\beqa}{\begin{eqnarray}}
\newcommand{\eeqa}{\end{eqnarray}}

\newcommand{\bn}{\hat{\bf n}}

\newcommand{\len}{\phi}
\newcommand{\TTest}{\widehat C_l^{\cmb\cmb}}
\newcommand{\TPest}{\widehat C_l^{\cmb\phi}}
\newcommand{\PPest}{\widehat C_l^{\phi\phi}}
\newcommand{\TT}{C_l^{\cmb\cmb}}
\newcommand{\TP}{C_l^{\cmb\phi}}
\newcommand{\PP}{C_l^{\phi\phi}}
\newcommand{\rest}{\widehat r_l}

\begin{document}
\twocolumn[\hsize\textwidth\columnwidth\hsize\csname
@twocolumnfalse\endcsname

\title{Can Cosmic Shear Shed Light on Low Cosmic Microwave
Background Multipoles?}
\author{Michael Kesden, Marc Kamionkowski, and Asantha Cooray}
\address{
Mail Code 130-33, California Institute of Technology, Pasadena,
California 91125}

\date{June 2003}

\maketitle


\begin{abstract}
  The lowest multipole moments of the cosmic microwave
  background (CMB) are smaller than expected for a
  scale-invariant power spectrum.  One possible explanation is a cutoff in the
  primordial power spectrum below a comoving scale of $k_c \simeq 5.0
  \times 10^{-4}$ Mpc$^{-1}$.  This would affect not only the
  CMB but also the cosmic-shear (CS)
  distortion of the CMB.  Such a cutoff increases significantly the
  cross-correlation between the large-angle CMB and cosmic-shear
  patterns.  The cross-correlation may be detectable at
  $> 2\sigma$ which, when combined with the low CMB moments,
  may tilt the balance between a $2\sigma$ result and a firm
  detection of a large-scale power-spectrum cutoff.  As an
  aside, we also note that the cutoff increases 
  the large-angle cross-correlation between the CMB and
  low-redshift tracers of the mass distribution.
\end{abstract}



]


One of the more intriguing results to come from the Wilkinson
Microwave Anisotropy Probe (WMAP) \cite{WMAPresults}
is confirmation of the absence of large-scale 
temperature correlations in the cosmic microwave background
(CMB), or equivalently, a suppression of power in the quadrupole
and octupole moments, found earlier by the Cosmic Background
Explorer \cite{cobe}.  A variety
of measures of the power spectrum---from the $l=4$ moment of the
CMB power spectrum, which probes wavelengths $\sim10^4$ Mpc, to
galaxy surveys and the Lyman-alpha forest, which probe down to 1--10
Mpc---show consistency with a scale-invariant spectrum of
primordial perturbations.  Thus, the suppression of the $l=2$
and $l=3$ moments of the CMB power spectrum come as a bit of a
surprise.

Is this simply a statistical fluke?  Or is something novel
occurring just beyond our observable cosmological horizon?
Possibilities include remnants of a pre-inflationary Universe, a
curvature scale just larger than the horizon, and/or exotic
inflation \cite{curv,limit}.  If there is indeed a
suppression of large-scale power, it occurs at distance scales
$\sim10^4$ Mpc \cite{limit}, larger than those typically probed
by galaxy surveys.  Future experiments to determine the lowest
moments of the CMB power spectrum are also of limited value
because current measurements are already dominated by cosmic
variance rather than instrumental noise.  Thus, although the
current evidence for new super-horizon physics is tantalizing,
the prospects for further testing it are limited.

In this paper we point out that there exists another probe of
the mass distribution on these largest distance scales.  Cosmic
shear (CS), weak gravitational lensing by density perturbations along the
line of sight, will produce identifiable distortions in the
temperature-polarization pattern of the CMB.  When observed,
these distortions map the gravitational potential projected
along a given line of sight.  Here we show that a power-spectrum
cutoff enhances significantly (roughly a factor of four) the
cross-correlation between the CMB and CS distortion of the CMB
on the largest scales.  This cross-correlation may be detectable
at the $> 2\sigma$ level and may thus provide a valuable
cross-check to the current $\sim2\sigma$ evidence for a dearth
of large-scale CMB power.  As an aside, we also show that the large-angle
cross-correlation between the CMB and low-redshift tracers of
large-scale structure \cite{critturok} is roughly doubled if the
large-scale cutoff is real.  Although recent detections \cite{recentdetections}
of this effect are at smaller scales than would be affected by a large-scale
cutoff, correlations on larger scales might be probed by future experiments.

Below we first discuss the large-scale CS power
spectra, as well as the cross-correlation of the CS
pattern with the CMB temperature pattern.  We then construct an
estimator for the cross-correlation, and show that it can
distinguish the cross-correlation with and without a cutoff at
roughly the $2\sigma$ level.  When combined with the already
suspiciously low $l=2$ and $l=3$ moments of the CMB power
spectrum, this finding may tilt the balance between a $2\sigma$
result and a $3\sigma$ discrepancy with scale invariance.

Perturbations in the matter density induce perturbations to the
gravitational potential $\Phi(\bfr,z)$ which then induce
temperature perturbations in the CMB through the Sachs-Wolfe
effect
\begin{equation} \label{E:ISWn}
     \cmb(\bn) = \frac{1}{3} \Phi(\bfr_0,z_0) - 2 \int_0^{r_0}
     \frac{d\Phi}{dr}(\bfr,z(r)) \, dr \, ,
\end{equation}
where $\bfr$ and $z$ are the physical comoving distance and
redshift, respectively, and the subscript 0 denotes these quantities at the
last-scattering surface.  The position vector $\bfr$ points in the direction
$\bn$ on the sky.  The potential at redshift $z$ can be
related to its present-day value with the linear-theory growth
factor $G(z)$ (normalized to unity today) through $\Phi(\bfr,z)
= (1+z) G(z)\Phi(\bfr,0)$.  The first term in
Eq. (\ref{E:ISWn}) comes from density perturbations at the
surface of last scatter, while the second term (the integrated
Sachs-Wolfe effect; ISW) comes from density perturbations along
the line of sight.

Relating the potential to
the matter perturbation through the Poisson equation, if the
three-dimensional matter power spectrum is $P(k)$ as
a function of wavenumber $k$, then the angular power spectrum
for temperature fluctuations is
\begin{equation} \label{E:ISWpow}
     C_l^{\cmb\cmb} \propto \int dk\,
     k^{-2} P(k) [\widetilde{\cmb_l}(k)]^2,
\end{equation}
as a function of multipole moment $l$, where
\begin{eqnarray} \label{E:cmbtil}
     \widetilde{\cmb_l}(k) &=& \frac{1}{3} (1+z_0) G(z_0)
     j_l(kr_0) \nonumber \\
     &-& 2 \int_0^{z_0} dz \left[ (1+z) G(z) \right]^{\prime}
     j_l\left(kr(z)\right) \, ,
\end{eqnarray}
and the prime denotes derivative with respect to redshift $z$.
The main contributions to the integral in Eq. (\ref{E:ISWpow})
come from wavenumbers $k$ near $10^{-4}$ Mpc$^{-1}$ (see, e.g., Fig. 7 in
Ref. \cite{KamSpe94}).

\begin{figure}[t]
\begin{center}
\epsfig{file=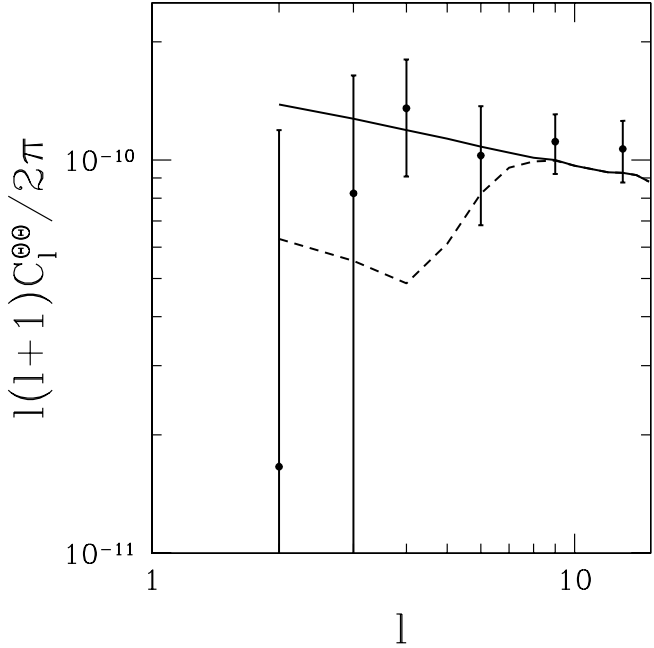,height=9.0cm,bbllx=15,bblly=140,bburx=580,bbury=710,clip=}
\caption{The CMB temperature power spectrum.  The solid curve is the power
     spectrum of Eq.~(\ref{E:ISWpow}) for the ISW effect without
     a cutoff.  The dashed curve has a cutoff in $P(k)$ below
     $k_c = 5.0 \times 10^{-4}$ Mpc$^{-1}$.  The binned error bars represent
     actual WMAP data.
}
\label{F:WMAP}
\end{center}
\end{figure}

If scale invariance holds out to super-horizon scales (as
predicted by the generic inflationary model), then the power
spectrum $P(k)$ at distance scales relevant for 
$l\lesssim10$ is simply $P(k) \propto k^n$, with $n$ near
unity.  This is certainly what the CMB data show at multipole
moments $l\geq4$ and it is consistent with determinations of the
power spectrum from the CMB and large-scale structure out to
scales as small as $\sim$few Mpc.  Thus, the observed suppression
of $C_2^{\cmb\cmb}$ and $C_3^{\cmb\cmb}$ shown in
Fig. \ref{F:WMAP} is a bit of a surprise.

The same potential perturbations $\Phi(\bfr,z)$ that contribute
to the Sachs-Wolfe effect also give rise to weak gravitational
lensing described by the projected potential,
 \begin{equation} \label{E:lpn}
     \phi(\bn) = -2 \int_0^{r_0} dr \frac{r_0 - r}{r_0 r} \Phi(\bfr,z(r)) \, .
\end{equation}
The angular power spectrum of the lensing potential is then
\begin{equation} \label{E:lpPS}
     C_l^{\phi\phi} \propto \int dk\,k^{-2} P(k)
     [\widetilde{\phi_l}(k)]^2 \, .
\end{equation}
In fact, the only difference
between this expression and its SW counterpart is
the replacement of $\widetilde{\cmb_l}(k)$ by
\begin{equation} \label{E:phitil}
     \widetilde{\phi_l}(k) = -2 \int_0^{z_0} \frac{c \, dz}{H(z)}
\frac{r_0 - r(z)}{r_0 r(z)} (1+z) G(z) j_l(kr(z)).
\end{equation}
The projected potential receives contributions from a wide
variety of distances, peaked at roughly half the comoving
distance to the surface of last scatter.  The lowest
multipole moments of the CS power spectrum come from
wavenumbers $k$ near $10^{-4}$ Mpc$^{-1}$.  Since the small-$k$ Fourier modes
of the potential that give rise to low-$l$ CS moments are the same as those
that give rise to the low-$l$ CMB moments, we anticipate that the CS power
spectrum should also reflect the suppression of large-scale power.
Evaluating these expressions numerically, however, we find that
the cutoff suppresses $C_2^{\phi\phi}$ by no more than
$\sim10\%$, too small to be detected.

However, the CMB and CS multipole moments are generated by the
same underlying potential fluctuations, and so there should be some
cross-correlation between the two.  And as we show, this
cross-correlation turns out to be increased significantly if
there is a cutoff.   The cross-correlation power spectrum
$C_l^{\cmb\phi}$ is 
\begin{equation} \label{E:cmblpCC}
     C_l^{\cmb\phi} \propto \int dk\, k^{-2} P(k)
     \widetilde{\cmb_l}(k) \widetilde{\phi_l}(k) \, .
\end{equation}
We can define a dimensionless cross-correlation coefficient,
$r_l \equiv (\TP)^2/\TT\PP$.  If $\widetilde \Theta_l(k)$ and $\widetilde
\phi_l(k)$ had precisely the same $k$ dependence, then the CMB
maps would be maximally correlated, $r_l=1$.  In this case, we
would be able to predict precisely that the CS spherical-harmonic
coefficients should be $\len_{lm}=(\TP/\TT) \cmb_{lm}$ in terms
of the measured temperature coefficients $\cmb_{lm}$.  Moreover, if $r_l$
were equal to unity, then a CS map might be used to
confirm the CMB measurements, but it would add no additional
statistically-independent information on the large-scale power
spectrum.

If, on the other hand, there was no overlap between $\widetilde
\Theta_l(k)$ and $\widetilde \phi_l(k)$ whatsoever, then there would
be no cross-correlation, $r_l=0$.  In this case, the CS
pattern could not confirm the CMB measurement, but it would
provide a statistically independent probe of the large-scale
power spectrum.

\begin{figure}[t]
\begin{center}
\epsfig{file=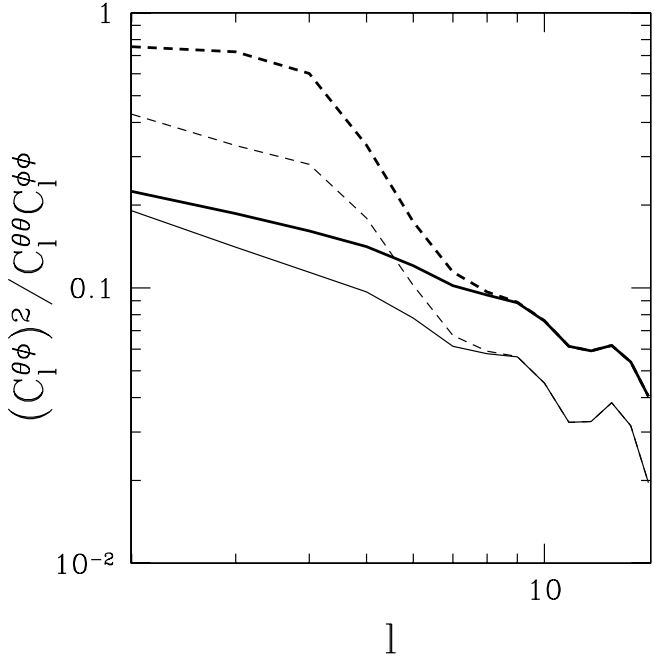,height=9.0cm,bbllx=15,bblly=140,bburx=580,bbury=710,clip=}
\caption{The dimensionless correlation $(\TP)^2/\TT\PP$ between maps of the CMB
     temperature and lensing potential.  The solid curve shows
     this correlation in the absence of a cutoff, while the
     dashed curve is for a cutoff $k_c = 5.0 \times 10^{-4}$
     Mpc$^{-1}$.  The upper, darker curves correspond to the lensing potential
     seen by sources at the CMB last-scattering surface, while the lower,
     lighter curves correspond to lensing sources at redshift $z = 1.0$.
}
\label{F:CC}
\end{center}
\end{figure}

Most generally, $0<r_l<1$, and the lensing spherical-harmonic
coefficients will be 
\begin{equation} \label{E:parcormm}
     \len_{lm} = (\TP/\TT) \cmb_{lm} + \left[\PP - (\TP)^2/\TT)
     \right]^{1/2} \zeta \, ,
\end{equation}
where $\zeta$ is a Gaussian random variable with zero mean and
unit variance; i.e., there is a correlated part determined by
the CMB pattern and an uncorrelated part.

Fig.~\ref{F:CC} shows our central result: the cross-correlation
coefficient for a scale-invariant spectrum and one in
which $P(k)=0$ for $k<k_c=5\times10^{-4}$ Mpc$^{-1}$.  The
dramatic increase
in the cross-correlation for the lowest $l$ in the presence of a
cutoff can be understood by examining the two terms of
Eq.~(\ref{E:ISWn}).  The first of these terms, generated at the
last-scattering
surface, is uncorrelated with the lensing potential; the second
term is a line-of-sight integral like the projected potential of
Eq.~(\ref{E:lpn}).  Since the contribution of the first term
comes from a larger distance from the observer than that of the
second term, correspondingly larger structures with lower
wavenumber $k$ will be projected onto the angular scale set by
the multipole moment $l$.  The lowest multipole moments will
correspond to structures at the last-scattering surface with $k
< k_c$, implying that in the presence of a cutoff only the
second term of Eq.~(\ref{E:ISWn}) will be nonvanishing for the
lowest multipole moments. Since it is only this term that is
correlated to the lensing potential, the dimensionless
cross-correlation will be significantly higher in the presence
of a cutoff.  It is important to note that this increase in the dimensionless
cross-correlation in the presence of a cutoff is an independent prediction
and not merely a consequence of the observed suppression of $\TT$ for low $l$.
If the CMB and CS multipole moments $\cmb_{lm}$ and $\len_{lm}$ were multiplied
by an $l$-dependent normalization to suppress power on large scales, $r_l$
itself would remain unaffected because it is dimensionless.  The independence
of this prediction allows estimates of $r_l$ from CMB and CS maps to constrain
$k_c$ with greater statistical significance than measurements of $\TT$ alone.

\begin{figure}[t]
\begin{center}
\epsfig{file=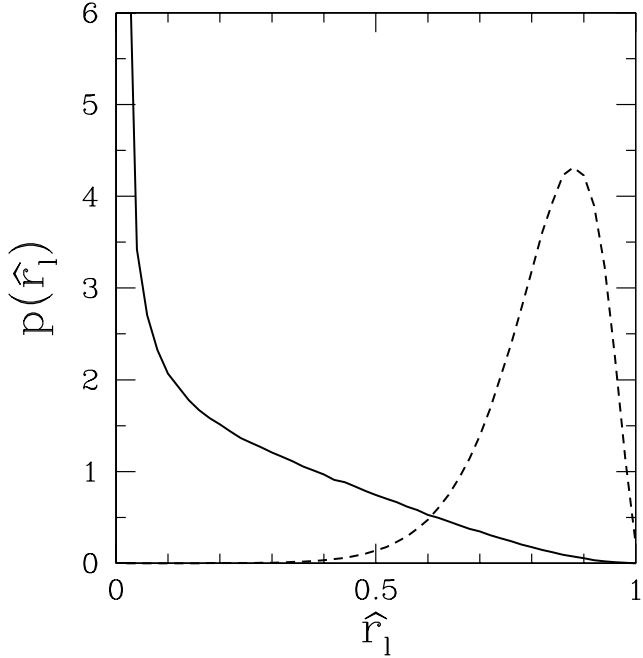,height=9.0cm,bbllx=15,bblly=140,bburx=580,bbury=710,clip=}
\caption{Probability distributions for $\rest$ for $l=3$ for the two models
described in the text; the solid curve corresponds to the model without a
cutoff in $P(k)$ while the dashed curve has a cutoff $k_c = 5.0 \times 10^{-4}$
Mpc$^{-1}$.}
\label{F:prob}
\end{center}
\end{figure}

We now determine how well measurements of $r_l$ can discriminate
between a model with a scale-invariant power spectrum and one with a cutoff
(i.e., $P(k)=0$ for $k<k_c= 5.0 \times 10^{-4}$ Mpc$^{-1}$).
Higher-order correlations in a high-resolution low-noise CMB
temperature-polarization map, can be used to construct
estimators \cite{lensing} for the projected potential and thus
estimators $\TPest$ and $\PPest$, in addition to those $\TTest$
for the temperature obtained already by WMAP.
An estimator $\rest \equiv (\TPest)^2/\TTest\PPest$ for $r_l$
can then be formed.  Although it is not an unbiased estimator,
it is sensitive to $r_l$
and converges to $r_l$ for $l\gg1$.  For a given
realization of the CMB and CS patterns, $r_l$ can be estimated independently
for each value of $l$.  We have calculated the probability
distributions for $\rest$ for each $l$ from many different Monte
Carlo realizations of the two models for $r_l$ described
above.  The CMB coefficients $\cmb_{lm}$ are set to the values consistent with
the WMAP power spectrum shown in Fig.~\ref{F:WMAP},
while the uncorrelated part of $\len_{lm}$ is determined for each
realization of the two models in accordance with Eq.~(\ref{E:parcormm}).
We assume the CS projected potential is reconstructed
from a full-sky CMB temperature-polarization map with $7'$
angular resolution and noise-equivalent temperature of 0.46
$\mu{\rm K} \sqrt{{\rm sec}}$.  The different predictions for
${\widehat r_3}$ for the two models are shown in
Fig.~\ref{F:prob}; the predictions for ${\widehat r_2}$ and
${\widehat r_4}$ are qualitatively similar while for ${\widehat
r_5}$ the two probability distributions begin to merge and for
${\widehat r_6}$ they are almost indistinguishable.

Assuming that the first model (no cutoff) is correct, we
calculated the fraction of realizations in which
the measured values of $\rest$ would lead us to conclude that they were more
likely drawn from the probability distributions of the second model.  This
occurs only 0.7\% of the time, implying that only 0.7\% of the time would
cosmic variance mislead us into thinking that a cutoff as large as $k_c = 5.0
\times 10^{-4}$ Mpc$^{-1}$ was favored over the no-cutoff model.  Since
the CMB coefficients $\cmb_{lm}$ are constrained to a single realization
consistent with WMAP in
both models, this measurement would be statistically independent of the low
observed CMB multipole moments themselves for the purpose of distinguishing
the two models.  It could thus increase the
$\sim2\sigma$ discrepancy of that measurement into a $>3\sigma$
detection of a large-scale cutoff in the power spectrum.

Finally, we consider the cross-correlation between the CMB and
low-redshift ($z\lesssim1$) tracers of large-scale structure
\cite{critturok} that several groups have already claimed to
detect \cite{recentdetections}.  To estimate the effect of a
cutoff, we have calculated the cross-correlation between the CMB
and CS of a hypothetical population of sources at redshift
$z=1$ for models with and without a cutoff.  As indicated by the
lighter curves in Fig.~\ref{F:CC}, a cutoff
boosts the cross-correlation coefficient for these low-redshift
sources by roughly a factor of two.  Similar results are
obtained for cross-correlation with the galaxy distribution.
Although recent detections
of the cross-correlation occur at smaller scales ($l \simeq 30$) than those
effected by the cutoff, future large-scale surveys could be sensitive to a
cutoff-induced enhancement.

The WMAP observations clearly support the $\Lambda$CDM concordance
model, but they do present a few tantalizing discrepancies.  Perhaps the most
intriguing is the sharp decrease in observed power at the lowest
multipole moments shown in Fig.~\ref{F:WMAP}.  A variety of fundamental causes
for this large-scale suppression can be modeled empirically by an effective
cutoff $k_c$ in the primordial power spectrum $P(k)$.  Though the WMAP team
found that only $0.15\%$ of simulated CMB maps had less power on large scales
\cite{WMAPresults}, it would be highly desirable to find corroborating evidence
to confirm that this observation is not merely a statistical anomaly.
One possibility is measurements of the polarization signal from nearby galaxy
clusters, which are proportional to their local CMB quadrupole moment
\cite{CooBau}.  We have proposed here that the cross-correlation $\TP$ could
provide additional
evidence.  Although the experimental requirements for
measurements of the CS distortions to the CMB are ambitious,
they are closely aligned with those for the CMBPOL experiment
that appears in NASA's roadmap.  Thus, this measurement, like
the effect it seeks to study, is on the horizon.

\acknowledgements
This work was supported in part by NASA NAG5-11985 and DoE
DE-FG03-92-ER40701.  Kesden acknowledges the support of an NSF
Graduate Fellowship.


\begin{thebibliography}{99}
\frenchspacing

\bibitem{WMAPresults} C. L. Bennett et al., astro-ph/0302207;
     D. N. Spergel et al., astro-ph/0302209. 

\bibitem{cobe} C. L. Bennett et al., Astrophys. J. Lett. {\bf
     464}, L1 (1996).

\bibitem{curv} G. Efstathiou, astro-ph/0303127; A. Blanchard et
     al., astro-ph/0304237; B. Feng and X. Zhang,
     astro-ph/0305020; M. Kawasaki and F. Takahashi,
     hep-ph/0305319; S. DeDeo, R. R. Caldwell, and P. J. Steinhardt, \PRD\
     {\bf 67}, 103509 (2003).

\bibitem{limit} C. R. Contaldi et al., astro-ph/0303636;
     E. Gaztanaga et al., astro-ph/0304178; G. Efstathiou,
     astro-ph/0306431.  J. M. Cline, P. Crotty, and
     J. Lesgourgues, astro-ph/0304558.


\bibitem{critturok} R. G. Crittenden and N. Turok,
     Phys. Rev. Lett. {\bf 76}, 575 (1996); M. Kamionkowski,
     Phys. Rev. D {\bf 54}, 4169 (1996); S. P. Bough,
     R. G. Crittenden, and N. G. Turok, New Astron. {\bf 3}, 275
     (1998); A. Kinkhabwala and M. Kamionkowski,
     Phys. Rev. Lett. {\bf 82}, 4172 (1999); U. Seljak and
     M. Zaldarriaga, Phys. Rev. D {\bf 60}, 043504 (1999).

\bibitem{recentdetections} S. P. Boughn and R. G. Crittenden,
     astro-ph/0305001; M. R. Nolta et al., astro-ph/0305097;
     P. Fosalba, E. Gaztanaga, and F. Castander,
     astro-ph/0307533; P. Fosalba and E. Gaztanaga,
     astro-ph/0305468; R. Scranton et al. (SDSS Collaboration),
     astro-ph/0307335; N. Afshordi, Y. Loh, and M. Strauss, astro-ph/0308260.

\bibitem{KamSpe94} M. Kamionkowski and D. N. Spergel, Astrophys.
     J. {\bf 432}, 7 (1994).
 
\bibitem{lensing}  U. Seljak and M. Zaldarriaga, \PRL\ {\bf 
     82}, 2636 (1999); W. Hu, \PRD\ {\bf 64}, 083005 (2001);
	W. Hu, \ApJL\ {\bf 557}, L79 (2001); W. Hu and
	T. Okamoto, \ApJ\ {\bf 574}, 566 (2002); M. Kesden,
	A. Cooray, and M. Kamionkowski, \PRL\ {\bf 89}, 011304
	(2002); L. Knox and Y.-S. Song, \PRL\ {\bf 89}, 011303;
	M. Kesden, A. Cooray, and M. Kamionkowski, \PRD\ {\bf
	67}, 123507 (2003); C. Hirata and U. Seljak, astro-ph/0306354.

\bibitem{CooBau} A. Cooray and D. Baumann, \PRD\ {\bf 67}, 063505 (2003);
  M. Kamionkowski and A. Loeb, \PRD\ {\bf 56}, 4511 (1997).

\end{thebibliography}
\end{document}